\documentclass[useAMS,usenatbib]{mn2e}
\usepackage{epsfig,amssymb}
\title[Identification of
sources of the highest energy EGRET
photons by correlation analysis]{Identification of
extragalactic
  sources of the highest energy EGRET
photons by correlation analysis}
\author[D. S. Gorbunov, P. G. Tinyakov, I. I. Tkachev
and S. V. Troitsky]
{D. S. Gorbunov$^{1}$,
P. G. Tinyakov$^{2,1}$,
I. I. Tkachev$^{3,1}$ and
S. V. Troitsky$^{1}$\thanks{E-mail: st@ms2.inr.ac.ru}\\
$^a$ Institute for Nuclear Research of the Russian Academy of
Sciences,
60th October Anniversary prosp. 7a, 117312, Moscow, Russia\\
$^b$Service de Physique
Th\'eorique, Universit\'e Libre de Bruxelles, CP225, blv.~du Triomphe, B-1050
Bruxelles, Belgium\\
$^c$ CERN Theory Division, CH-1211 Geneva 23, Switzerland
}
\begin{document}
%\date{in original form 2005 May 30}
\date{}
\pagerange{\pageref{firstpage}--\pageref{lastpage}} \pubyear{2005}
\maketitle
\label{firstpage}
\begin{abstract}
We found significant correlations between the arrival directions of
the highest energy photons ($E_\gamma>10$~GeV) observed by EGRET and
positions of the BL Lac type objects (BL Lacs). The observed
correlations imply that not less than three per cent of extragalactic
photons at these energies originate from BL Lacs. Some of the correlating
BL Lacs have no counterparts in the EGRET source catalog, i.e. do not
coincide with strong emitters of gamma-rays at lower energy. The study of
correlating BL Lacs suggests that they may form a subset which is
statistically different from the total BL Lac catalog; we argue that they
are prominent candidates for TeV gamma-ray sources. Our results
demonstrate that the analysis of positional correlations is a powerful
approach indispensable in cases when low statistics limits or even
prohibits the standard case-by-case identification.
\end{abstract}


\section{Introduction}
Identification of point sources observed at different wavelengths is a
standard problem in astronomy. This problem becomes very difficult at
high energies, when the angular resolution decreases and a mere
positional coincidence becomes insufficient for the identification (an
example is the EGRET catalog of point sources \citep{3EG} more than a
half of which have no optical/radio counterparts). The photon flux
also decreases with energy. Already at EGRET energies the separation
of point sources from the diffuse background becomes a challenging
problem by itself. At even higher energies, the flux becomes so low
that it is more appropriate to think of the data as a collection of
individual photons. In this situation, the standard method which
consists in finding local excesses of flux and identifying these
excesses with astrophysical objects becomes inadequate.

In this paper we argue that in the case of low flux the problem of
identification can be approached by an alternative method based on the
statistical analysis of correlations between individual photons and a
given {\em catalog} of astrophysical objects. The advantage of this
method is that it may give positive identification even in the case
when the average number of photons from a source is much less than
one. The price to pay is the statistical character of the obtained
information: one may establish with certainty that the catalog
contains actual sources, but may, in general, not be able to identify
them individually, nor estimate the luminosity of a given source. This
method has been previously applied to the case of ultra-high energy
cosmic rays by \citet{BL1,BL:GMF,BL:EGRET,BL:HiRes,comparative}.

Here we apply this method of identification to the catalog of
individual photons with energies $E_\gamma > 10\mbox{~GeV}$ published
recently by \citet{EGRET:highest}. We demonstrate that the arrival
directions of these photons correlate with positions of optically bright BL
Lacertae type objects (BL Lacs), that is the BL Lacs are emitters of
hard gamma-rays. We then analyse the emission and spectral properties
of the correlating objects at different wavelengths and discuss
possible implications of this analysis, in particular for the prospects of
TeV observations.

\section{Catalogs of candidate sources and gamma-rays}
\label{sec:sources}
Blasars, and the BL Lacs in particular, are the active galaxies with
jets pointing to the Earth.  Many of them are known to be strong
gamma-ray emitters (see, e.g.,
\citet{3EG,strong-emitters1,strong-emitters2}.  They are believed to
host powerful astrophysical accelerators of very energetic particles.

Current understanding of physical propeties of BL Lacs
is far from being complete. The standard approach
suggests the two-bump broadband spectral energy distribution (SED;
see, e.g., \citet{PadovaniGiommi95,Fossatti_et_al98}). The
lower-frequency peak, whose position varies from the infrared
(low-energy-peaked BL Lacs, or LBL) to the X-ray (high-energy-peaked,
or HBL) band, is often well-measured and is believed to be caused by
the synchrotron radiation of relativistic electrons.  The second bump
is probably due to the synchrotron photons scattering off the same
relativistic electrons
(self-Compton, SC); its position should be correlated with that of the
first bump and varies from MeV (LBL) to TeV (HBL) gamma-rays. Due to
a scarcity of the gamma-ray data, the SC bumps are generally studied
much worse than the synchrotron ones.

While the two-bump SED is inherent to the electron-powered jets,
strong gamma-ray emission from BL Lacs is expected also in other
models, e.g.\ in the `proton blazar' model \citep{proton-blazars}
where it inevitably accompanies the acceleration of protons. Not
surprisingly, many of blasars have been identified as EGRET sources
\citep{3EG,strong-emitters1,strong-emitters2}.
%, determined as local $4
%\sigma$ excesses of flux over the diffuse background at $E>100$~MeV as
%observed by the EGRET instrument.
There are no reasons for the gamma-ray spectra of blasars to have a
cutoff at the EGRET energies; at least some of the objects are likely
emitters of photons at higher frequencies. In a few cases, this has
been confirmed by the TeV observations.

The Table of BL Lac's (Table II) in the catalog of quasars and active
galactic nuclei \citep{Veron2003} consists of objects with different
spectral properties which are divided into three classes --
confirmed (`BL'), high-polarization (`HP'), and probable/possible (`BL?')
BL Lacs. This division is made according to several criteria
(some of which are discussed by \citet{veron})
and may reflect important differences in physical properties of the
objects. In our correlation analysis we test these three subclasses
separately. Each of them we divide in addition into optically bright
($V<18$ mag) and dim ($V\geq 18$ mag) parts. In this
way we obtain three subsamples of optically bright BL Lac's:

(1) The set of all confirmed `BL' type objects with the visual
magnitude $V< 18$ mag.  This set of BL Lac's contains 178
objects.

(2) The set of all confirmed `HP' type objects with the visual
magnitude $V<18$ mag. This set of BL Lacs contains 47 objects.

(3) The rest of the objects listed in the Table II of the catalog
\citep{Veron2003} with the visual magnitude $V<18$ mag. This set
consisting of bright unconfirmed BL Lac's contains 81 object.

The catalog of EGRET gamma-rays of the highest energies
($E_\gamma>10$~GeV) contains 1506 events \citep{EGRET:highest}.  To
suppress the background of Galactic gamma-rays we make a cut on the
Galactic latitude, $b>10^\circ$. This reduces the number of events
down to 613.

\section{Procedure}
\label{sec:procedure}

Our analysis is based on the calculation of the angular correlation
function as described by \citet{BL1}. The statistical significance of
correlation is estimated by testing the hypothesis that the highest
energy gamma-rays observed by EGRET and candidate sources are {\em
uncorrelated}. This is done as follows. For a given set of sources and
the angle $\delta$, we count the number of pairs {\em source -- gamma
ray} separated by the angular distance less or equal to $\delta$, thus
obtaining the {\em data count}. We then replace the real data by a
randomly generated Monte-Carlo set of gamma-rays and calculate the
number of pairs in the same way, thus obtaining the {\em Monte-Carlo
count}. We repeat the latter procedure many times calling {\em
successful } those tries when the Monte-Carlo count equals or exceeds
the data count. The number of successful tries divided by the total
number of tries gives the probability $P(\delta)$ that the excess
in the data count occured by chance.
The smaller is this probability, the stronger (more significant) is
the correlation. The validity of this straightforward approach does
not depend~\citep{cuts-and-penalties} on the completeness of the
catalog of the candidate sources provided simulated sets of events
correctly represent the detector exposure.

The Monte-Carlo events are drawn from the pool of events generated
according to the EGRET exposure map (available at {\tt
ftp://cossc.gsfc.nasa.gov/compton/data/} {\tt
egret/high\_level/combined\_data/}).  The latter depends on energy; we
adopt the map relevant for the highest energy range $4~{\rm GeV}
\lesssim E_\gamma \lesssim 10~{\rm GeV}$.  In our case the energy
range is even higher; however, the corresponding exposure map is
not available. We expect that this does not significantly influence
our results.

The significance of correlations is determined by the probability
$P(\delta)$ evaluated at the optimum value of $\delta$ which can
be obtained by Monte-Carlo simulations \citep{BL1,BL:HiRes}.
This optimum value is  usually close to the detector
angular resolution. The EGRET
detector has been carefully calibrated by \citet{EGRET-resolution};
its angular resolution depends on both the energy and the inclination
angle.  Averaged angular dispersion contains a narrow component and a
wide-angle tail and can be fitted at a given energy by four
Gaussians.
The radius of a circle containing
67\% of the gamma-rays depends on energy as
follows~\citep{EGRET-resolution},\footnote{The EGRET detector has
been calibrated at energies $E_\gamma\leq10$~GeV. In what follows we
assume that Eq.~(\ref{angular-resolution}) is valid at least up to
$E_\gamma\sim30$~GeV.}
\begin{equation}
\label{angular-resolution}
\delta_{67} (E_\gamma)\leq 0.50^\circ\left(\frac{10~{\rm
GeV}}{E_\gamma}\right)^{0.534}
\end{equation}
which gives an estimate for the angle $\delta$. Because of the
complexity of the EGRET angular resolution we do not fix $\delta$ by
the Monte-Carlo simulation. Instead, we follow an alternative approach
which consists in choosing the optimum bin size $\delta$ from the data
and correcting the corresponding significance by the penalty factor
\citep{BL1,cuts-and-penalties,Finley}. We will see in the next section
that this approach allows for a simple estimate of significance.

\section{Results}

\subsection{Positional correlations}
\label{sec:results}

\begin{figure}
\begin{center}
\includegraphics[width=8.4cm]{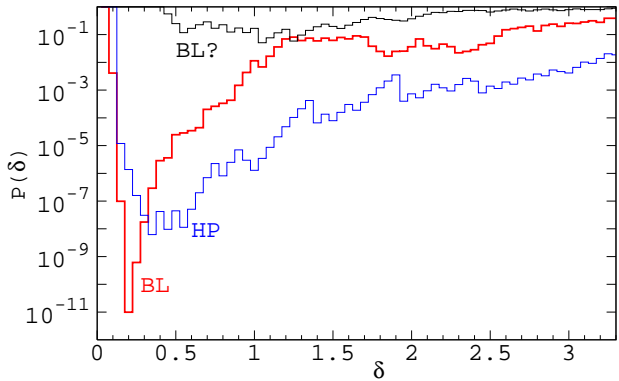}
\caption{$P(\delta)$ for the sets of bright ($V<18$mag) `BL',
`HP' and unconfirmed BL Lacs  (`BL?')  .
\label{fig:correlation}
}
\end{center}
\end{figure}

In Fig.\ref{fig:correlation} we present the probabilities $P(\delta)$ for all
three sets of optically bright objects. Sets (1) and (2) exhibit strong
correlations with the EGRET gamma-rays at separation angles
compatible with Eq.~(\ref{angular-resolution}), while correlations with the
set of unconfirmed BL Lacs are absent.

%%%%%%%%%%%%%%%%%%%%%%%%%%%%%%%%%%%%%%%%%%%%%%%%%%%%%%%%%%%%%%%%%%%%%%%%%%%%
%%%%%%%%%%%%%%%%%%%%%%%%%%%%%%%%%%%%%%%%%%%%%%%%%%%%%%%%%%%%%%%%%%%%%%%%%%%%

For the set (1), the minimum   value of $P(\delta)$ is $P
\approx 10^{-11}$, and there are 10 events which contribute to correlations
in the minimum (0.37 events expected as background from random
coincidences). For the set (2), the minimum  value is $P =
6.2\times 10^{-9}$ with 7 events contributing to
correlations (0.23 events expected from random
coincidences).\footnote{We report the probability calculated
from the data count and average Monte-Carlo count assuming the Poisson
distribution. This is a sufficiently good approximation at small
angular scales. The direct calculation of probabilities below
$10^{-5}$ is not feasible.
}

To obtain the significance of correlations one may take the lowest
value of $P(\delta)$ and multiply it by the penalty factor calculated
as described by \cite{BL1,cuts-and-penalties,Finley}.  This factor is
just the number of statistically independent ``attempts'' to find the
lowest probability. In the case at hand it can be replaced by a
conservative upper bound: the penalty factor for variation of a
quantity cannot exceed the number of steps, that is for $\delta $
varying between $0^\circ$ and $3^\circ$ in steps of $0.05^\circ$, the
penalty factor is $\le 60$. Multiplying the minimum probabilities in
Fig.~\ref{fig:correlation} by 60 clearly would not affect our
conclusions.

The minimum of $P(\delta)$ for BLs is at $\delta_{\rm min}^{BL} =
0.2^\circ$, while the minimum for HPs is at $\delta_{\rm min}^{HP} =
0.35^\circ$. More compact clustering of gamma-rays around BLs as comapared
to HPs can be explained by the difference in the typical energy of
correlating gamma-rays (here and in what follows we count the event as
correlating if it is at angular separation $\delta \leq \delta_{\rm
  min}$ for the respective set). As can be seen from Table~1, energies
of photons which correlate with BLs are systematically higher as
compared to those associated with HPs. It follows from
Eq.~(\ref{angular-resolution}) that the corresonding average angular
resolutions are $\langle \delta_{67}^{\rm (BL)}\rangle =0.34^\circ$,
$\langle \delta_{67}^{\rm (HP)}\rangle =0.46^\circ$, that could
explain the observed hierarchy $\delta_{\rm min}^{(1)}<\delta_{\rm
  min}^{(2)}$.
%%%%%%%%%%%%%%%%%%%%%%%%%%%%%%%%%%%%%%%%%%%%%%%%%%%%%%%%%%%%%%%%%%%%%%%%%%%%
\begin{figure}
\begin{center}
\includegraphics[width=8.3cm]{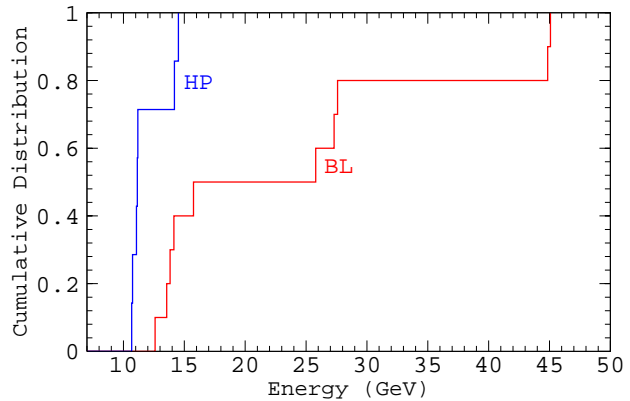}
\caption{Cumulative distribution of energies of correlating
gamma-rays for the sets of `BL' and `HP' type objects.
\label{fig:KS-energy}
}
\end{center}
\end{figure}
%%%%%%%%%%%%%%%%%%%%%%%%%%%%%%%%%%%%%%%%%%%%%%%%%%%%%%%%%%%%%%%%%%%%%%%%%%%%

Cumulative distributions of energies of gamma-rays correlating with BL
and HP are shown in Fig.~\ref{fig:KS-energy}.  The Kolmogorov--Smirnov
(KS) test gives $P$=1.4\% for the probability that both sets of
photons are drawn from one and the same distribution. The significance
is not high, so the systematic difference in energies could have
occured by chance. Nevertheless it gives a hint that `BL' and `HP'
objects may have physically different spectra of gamma-rays. This
issue cannot be elaborated further, in particular, because of the lack
of knowledge of the distances to the objects in
Table~\ref{Table-with-sources}.  Note that PKS 0208-512 has a redshift
$z$=1 implying that the observed photon had two times higher energy at
the source.
\begin{table}
\begin{tabular}{|c|c|c|c|c|c||c|}
\hline
name & t & Id &  $z$ &  $V$ & $F_{\rm 5}$ & $E_\gamma$  \\
\hline
\hskip -0.2cm PKS 2005-489  & \hskip -0.1cm BL \hskip -0.1cm & T  & 0.071& 12.8 & 1.19 & 13.8 \\
\hskip -0.2cm ON 231        & \hskip -0.1cm BL \hskip -0.1cm & \hskip -0.2cm E,G\hskip -0.2cm & 0.102& 16.1 & 0.72 & 27.3 \\
\hskip -0.2cm TXS 1914-194  & \hskip -0.1cm BL \hskip -0.1cm &    & 0.137& 15.3 & 0.41 & 27.6 \\
\hskip -0.2cm TXS 0506+056  & \hskip -0.1cm BL \hskip -0.1cm & G  &  ?   & 16.0 & 1.03 & 44.9 \\
\hskip -0.2cm              & \hskip -0.1cm \hskip -0.1cm   &    &      &      &      & 45.1 \\
\hskip -0.2cm RBS 76        & \hskip -0.1cm BL\hskip -0.1cm &    &  ?   & 16.3 & ?    & 25.8 \\
\hskip -0.2cm IVS B0621+446 & \hskip -0.1cm BL \hskip -0.1cm &    &  ?   & 16.8 & 0.37 & 12.6 \\
\hskip -0.2cm RGB J0806+595 & \hskip -0.1cm BL \hskip -0.1cm &    &  ?   & 17.2 & 0.04 & 13.6 \\
\hskip -0.2cm 3EG J0433+2908& \hskip -0.1cm BL \hskip -0.1cm & \hskip -0.2cm E,G\hskip -0.2cm &  ?   & 17.8 & 0.48 & 14.1 \\
\hskip -0.2cm              &  \hskip -0.1cm \hskip -0.1cm  &    &      &      &      & 15.7 \\
\hskip -0.2cm Mrk 421       & \hskip -0.1cm HP \hskip -0.1cm & \hskip -0.3cm E,G,T \hskip -0.3cm
& 0.031& 12.9 & 0.70 & 14.2 \\
\hskip -0.2cm PKS 2155-304  & \hskip -0.1cm HP \hskip -0.1cm & \hskip -0.2cm E,T\hskip -0.2cm & 0.116& 13.1 & 0.41 & 11.1 \\
\hskip -0.2cm              & \hskip -0.1cm \hskip -0.1cm    &     &      &      &      & 11.1 \\
\hskip -0.2cm              &  \hskip -0.1cm \hskip -0.1cm   &     &      &      &      & 11.2 \\
\hskip -0.2cm TXS 1215+303  & \hskip -0.1cm HP \hskip -0.1cm &     & 0.237& 15.6 & 0.42 & 10.7 \\
\hskip -0.2cm PKS 0208-512  & \hskip -0.1cm HP \hskip -0.1cm & \hskip -0.2cm E,G\hskip -0.2cm & 1.003& 16.9 & 3.21 & 10.7 \\
\hskip -0.2cm TXS 0912+297  & \hskip -0.1cm HP \hskip -0.1cm &     &  ?   & 16.4 & 0.20 & 14.5 \\
\hline
\end{tabular}
\caption{
\label{Table-with-sources}
BL Lac's from the samples (1), (2)
('BL' and 'HP' in column 't') and correlating gamma-ray events.  In
the column `Id', `E' indicates that the object is an EGRET
source~\citep{3EG} (positional identification of the corresponding
events with the 3EG sources has been claimed by
\citet{EGRET:highest}), `G' -- that it is a GeV source~\citep{GeV} and
`T' -- that it is a TeV source.  The visual magnitudes are presented
in column `$V$', the redshifts are presented in column `$z$' (the
question mark througout the table indicates that the value is
unknown); the column `$F_5$' presents the radio-flux at 5~GHz in Jy
($V$, $z$ and $F_5$ are taken from \citet{Veron2003}). Note that
3EG~J0433+2908 and TXS~0506+056 correlate with two gamma-rays each,
while PKS~2155-304 correlates with a triplet.
}
\end{table}

Optically faint objects, $V>18$ mag, of all three types (BL, HP
and unconfirmed BL Lac's) do not correlate with gamma-rays.  For
these subsets $P(\delta)\gtrsim 10\%$ in the range of $\delta$ compatible
with Eq.~(\ref{angular-resolution}).

\subsection{Physical properties of correlating BL Lacs}
\label{sec:properties}

The important question is which observational and/or intrinsic
properties may distinguish efficient high-energy gam\-ma-ray emitters.
To systematically study this issue we carry out the KS test for
compatibility of the distributions of correlating (without any cut on
$V$) and of all objects of the same type (BL, HP) with respect to
various parameters. Unlike correlations discussed in the previous
section, this study may be affected by incompleteness of the catalog
\citep{Veron2003}, so the results of this section should be
interpreted with care.

Some of the correlating BL Lacs are emitters of multiple photons, see
Table~1. It is possible to treat such objects in two ways. In the
first approach multiple emitters are treated on equal footing with the
rest of correlating BL Lacs. In the second approach multiple emitters
are given weight which equals the number of photons observed from them.  For the
sake of completeness we present the results of both approaches.

\begin{figure}
\begin{center}
\includegraphics[width=8.4cm]{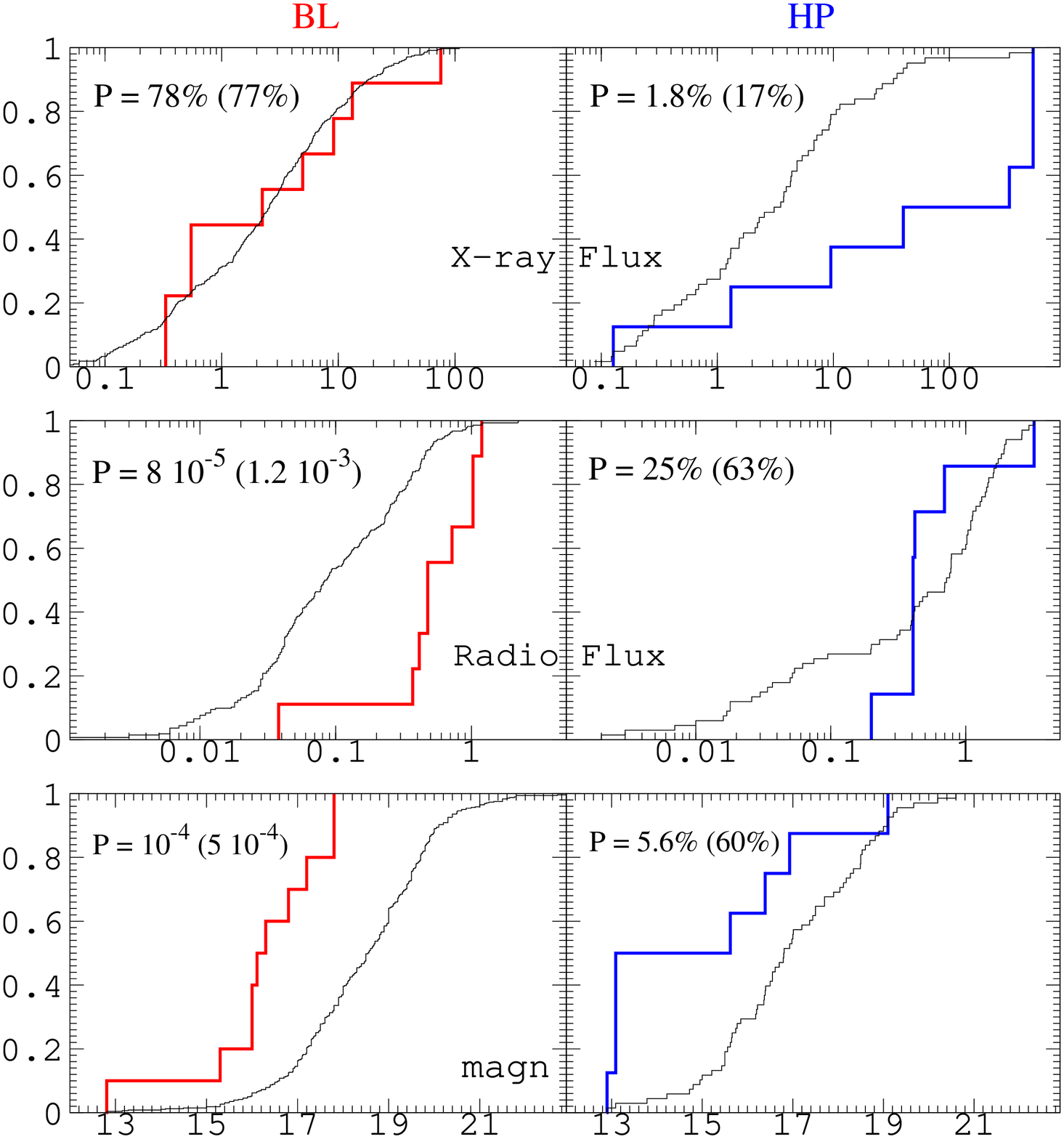}
\caption{Cumulative distributions of observational characteristics
for correlating (thick lines) and all (thin lines) objects of the types
`BL' (left panels) and `HP' (right panels). Upper panels show
distributions of X-ray fluxes ($10^{-12} {\rm mW/m^2}$), middle panels
correspond to radio flux (Jy) and bottom panels show apparent optical
$V$-magnitudes. ${\rm P}$ corresponds to the Kolmogorov-Smirnov
probability to obtain the distribution of correlating objects as a
statistical fluctuation of the distribution of all objects.
\label{figMulti_Indx4}}
\end{center}
\end{figure}

In Fig.  \ref{figMulti_Indx4} we compare the cumulative distributions
of visual magnitudes, radio-fluxes at 5~GHz and X-ray
fluxes\footnote{The catalog \citep{Veron2003} contains visual
magnitudes and radio-fluxes; X-ray fluxes are taken from the HEASARC
database ({\tt http://heasarc.gsfc.nasa.gov}) or calculated based on
the count rates given there.  Note that the emission and spectra of BL
Lacs vary strongly with time, so this study should be considered as
indicative only. } at 1~keV for correlating and all objects of types
`BL' and `HP'. Displayed curves correspond to the second approach when
multiple emitters are weighted according to the number of observed
photons. This allows to see positions of multiple emitters within
relevant distributions. The results of the KS test for the first approach
are given in parentheses.
%%%%%%%%%%%%%%%%%%%%%%%%%%%%%%%%%%%%%%%%%%%%%%%%%%%%%%%%%%%%%%%%%%%%%%%%%%%%
%%%%%%%%%%%%%%%%%%%%%%%%%%%%%%%%%%%%%%%%%%%%%%%%%%%%%%%%%%%%%%%%%%%%%%%%%%%%

\begin{figure}
\begin{center}
\includegraphics[width=8.4cm]{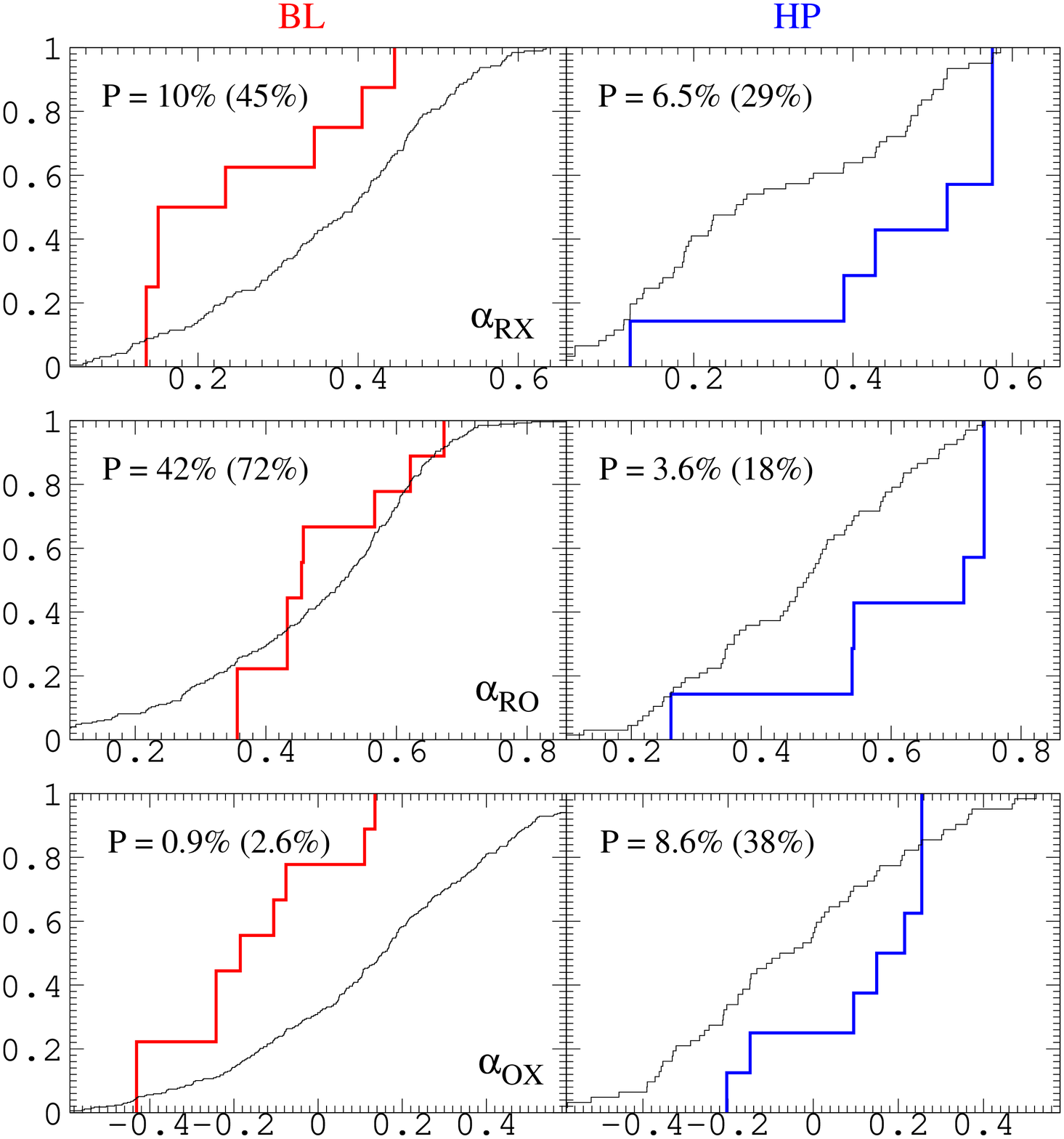}
\caption{Cumulative distributions of intrinsic spectral properties for
correlating (thick lines) and all (thin lines) objects of the types `BL'
(left panels) and `HP' (right panels). Upper, middle and bottom panels show
distributions of $\alpha_{RX}$, $\alpha_{RO}$ and $\alpha_{OX}$
correspondingly.
 ${\rm P}$ corresponds to the Kolmogorov-Smirnov probability to
obtain the distribution of correlating objects as a statistical
fluctuation of the distribution of all objects.
\label{spectral-indices}
}
\end{center}
\end{figure}

In Fig.~\ref{spectral-indices} we compare the cumulitive distributions
with respect to spectral indices.  The radio-to-optic $\alpha_{RO}$,
radio-to-X-rays $\alpha_{RX}$ and optic-to-X-rays $\alpha_{OX}$
indices are defined as
$\alpha_{AB}=\lg(\nu_A F_A/\nu_B F_B)/\lg(\nu_A/\nu_B)$,
where $F_A$ is the fluency at a frequency $\nu_A$.  These parameters
reflect intrinsic properties of the objects and are important for
understanding the acceleration mechanism operating inside the sources.
In particular, $\alpha _{OX}$ effectively measures the position of the
synchrotron bump in the blazar's SED.

%%%%%%%%%%%%%%%%%%%%%%%%%%%%%%%%%%%%%%%%%%%%%%%%%%%%%%%%%%%%%%%%%%%%%%%%%%%%
%%%%%%%%%%%%%%%%%%%%%%%%%%%%%%%%%%%%%%%%%%%%%%%%%%%%%%%%%%%%%%%%%%%%%%%%%%%%

Important physical information is contained also in the gamma-ray SEDs of
the objects, which are available for the EGRET sources only (two objects
of the BL type and three HPs). The EGRET spectral indices of these five
objects are unusually small, indicating higher fluency at higher energies.
This is fully consistent with identification of them as the sources of
$E_\gamma >10$~GeV photons.

The sources of the `HP' class have spectral indices $\alpha _{OX}\gtrsim
0$ indicating that most of them are the HBLs with the synchrotron bump in
X-rays. The IC bump is then expected at TeV energies, in consistency
both with the EGRET spectral indices (for the 3EG sources) and with our
evidence for energetic photons for the rest of the sample. Two of the
sources have indeed been already detected in TeV.

Much more interesting trends can be seen in the `BL' sample. Most of
these objects have significantly negative $\alpha _{OX}$ and are of the
LBL type, which may be also seen directly from the strong dominance of the
optical and radio over X-ray emission (see Fig.~\ref{figMulti_Indx4}),
suggesting the synchrotron bump in
the infrared. This usually corresponds to the sub-GeV IC bump, but the
possible detection of $E_\gamma >10$~GeV photons from them (and, in both
available cases, the EGRET spectral indices) indicate rising spectra at
GeV energies (cf.~the SED of ON~231 in \citet{Ghiss}).
Taken at face value, this fact has two immediate consequences:
firstly, these objects are good candidates for the detection in TeV (up to
now, only one of the eight has been detected); secondly, the two-bump
electron-blazar model may not work, in its classical form, for their SEDs.
The latter fact could be explained, for instance, by a special geometry of
the source~\citep{Bednarek} or by a more complicated gamma-ray-emission
model. Clearly, the evidence is insufficient to claim that the eight
sources considered here represent a new class (e.g., `proton blazars').
However, one may note that it is the `BL' objects brighter than 18\,mag
which correlate with ultra-high-energy cosmic rays\footnote{This
statement relates to the whole sample of 178 objects. Of eight
correlating BLs in Tab.~\ref{Table-with-sources}, only one falls in the
errorbox of the arrival directions of cosmic rays -- 3EG~J0433+2908
coincides with an AGASA-detected doublet.}, and the acceleration of
protons should take place in some of them, inevitably accompanied by the
emission of energetic gamma-rays. This tempting conjecture awaits further
studies with the existing (EGRET) and upcoming (GLAST) gamma-ray data.

Another interesting feature is related to the doublets and the triplet of
gamma-rays correlating with BL Lac's. All of them are
monochromatic, see Table~1. This may hint at the details of the
accelarating mechanism in the sources  or ambient matter density and
magnetic field strength close to the source.

As we have already pointed out, the small number of correlating
photons and incompleteness of the original catalog \citep{Veron2003}
prevent one from making definite conclusions on the basis of the
analysis performed in this subsection.
However, our results suggest
that the BL Lacs
which are bright in radio and optical band may often emit energetic
gamma-rays.
Together with
the fact that correlating BL's on average emit more energetic photons (in
$E_\gamma  > 10$ GeV band, see Fig.~\ref{fig:KS-energy}), this may
indicate
a deviation from the conventional two-bump SED model, possibly related to
the acceleration of cosmic rays.

\section{Conclusions}
\label{sec:concl}
The statistical analysis we performed in this paper reveals that the
arrival directions of the high-energy EGRET photons coincide with
positions of BL Lacs ('BL' and 'HP' objects) far too often to be
explained by chance: out of 10 coincidences observed for BLs only 0.37
(in average) would be expected if the photons and BLs were
uncorrelated. Thus, our analysis has established with certainty that
the BL catalog contains sources of gamma-rays. Moreover, nearly all
correlating photons are due to sources, and therefore all correlating
BLs are likely to be actual emitters (the same concerns HP objects).
On the other hand, one may check that there is no significant
clustering in the set of highest-energy EGRET photons, so the standard
methods of identification would be inconclusive.

The fact of correlations is not the only conclusion which statistical
methods are able to establish. As has been shown in sect.~4.2, they
allow to address a question of which physical property characterizes
the actual emitters. This, however, requires the completeness of the
catalog of candidate sources, as well as better statistics.

The authors are indebted to  M. Sazhin for
useful comments and to S. Sibiryakov for his
encouraging interest. The work was supported in part by the grant of the President
of the Russian Federation NS-2184.2003.2, by the INTAS grant
03-51-5112 (D.G.,  P.T.  and S.T.), by the grants of the Russian
Science Support Foundation and of the
"Dynasty" foundation (awarded by the Scientific board of ICFPM; D.G. and
S.T.), by IISN, Belgian
Science Policy (under contract IAP V/27; P.T.).
D.G. thanks CERN TH Division, where the scientific part of
 the work has been accomplished, for hospitality.

%%%%%%%%%%%%%%%%%%%%%%%%%%%%%%%%%%%%%%%%%%%%%%%%%%%%%%%%%%%%%%%%%%%%%%%%%

\bsp

\label{lastpage}


\begin{thebibliography}{99}

\bibitem[\protect\citeauthoryear{Bednarek}{1998}]{Bednarek}
 Bednarek W., 1998,
  %``Inverse Compton scattering model for gamma-ray production in Mev blazars,''
  MNRAS, 294, 439
  %[arXiv:astro-ph/9711188].
  %%CITATION = ASTRO-PH 9711188;%%

\bibitem[\protect\citeauthoryear{Finley \& Westerhoff}{2004}]{Finley}
Finley C.B., Westerhoff S, 2004,
  %``On the evidence for clustering in the arrival directions of AGASA's
  %ultrahigh energy cosmic rays,''
  Astropart.\ Phys., 21, 359
 % [arXiv:astro-ph/0309159].
  %%CITATION = ASTRO-PH 0309159;%%

\bibitem[\protect\citeauthoryear{Fossati et al.}{1998}]{Fossatti_et_al98}
Fossati G., Maraschi L., Celotti A., Comastri A., Ghisellini G., 1998,
  %``A Unifying View of the Spectral Energy Distributions of Blazars,''
MNRAS, 299, 433
  %arXiv:astro-ph/9804103.
  %%CITATION = ASTRO-PH 9804103;%%

\bibitem[\protect\citeauthoryear{Ghisellini}{2004}]{Ghiss}
Ghisellini G., 2004,
  %``The high energy view of blazars,''
  Nucl.\ Phys.\ Proc.\ Suppl., 132, 76
  %[arXiv:astro-ph/0308526].
  %%CITATION = ASTRO-PH 0308526;%%

\bibitem[\protect\citeauthoryear{Gorbunov \& Troitsky}{2004}]{comparative}
  Gorbunov D.S., Troitsky S.V., 2005,
  %``A comparative study of correlations between arrival directions of
  %ultra-high-energy cosmic rays and positions of their potential astrophysical
  %sources,''
  Astropart. Phys., 23, 175
 % [arXiv:astro-ph/0410741].
  %%CITATION = ASTRO-PH 0410741;%%

\bibitem[\protect\citeauthoryear{Gorbunov et al.}{2002}]{BL:EGRET}
Gorbunov D.S., Tinyakov P.G., Tkachev I.I, Troitsky S.V., 2002,
%``Evidence for a connection between
%gamma-ray and highest-energy cosmic  ray emissions by BL Lacs,''
ApJ, 577, L93
%%CITATION = ASTRO-PH 0204360;%%

\bibitem[\protect\citeauthoryear{Gorbunov et al.}{2004}]{BL:HiRes}
  Gorbunov D.S., Tinyakov P.G., Tkachev I.I, Troitsky S.V., 2004,
%, P.~G.~Tinyakov, I.~I.~Tkachev and S.~V.~Troitsky,
%``Testing the correlations between ultra-high-energy cosmic rays and BL Lac
  %type objects with HiRes stereoscopic data,''
  JETP Lett., 80, 145
%  [arXiv:astro-ph/0406654].
  %%CITATION = ASTRO-PH 0406654;%%

\bibitem[\protect\citeauthoryear{Hartman et al.}{1999}]{3EG}
Hartman R.C. et al., 1999, ApJS, 123, 79

\bibitem[\protect\citeauthoryear{Lamb \& Macomb}{1997}]{GeV}
Lamb R.C., Macomb D.J., 1997,
%``Point Sources Of Gev Gamma Rays,''
ApJ, 488, 872
%%CITATION = ASJOA,488,872;%%

\bibitem[\protect\citeauthoryear{Mannheim}{1993}]{proton-blazars}
Mannheim K., 1993,
  %``The Proton blazar,''
  A\&A,  269, 67
  %[arXiv:astro-ph/9302006].
  %%CITATION = ASTRO-PH 9302006;%%

\bibitem[\protect\citeauthoryear{Mattox et al.}{1997}]{strong-emitters2}
Mattox J.R., Schachter J., Molnar L., Hartman R.C., Patnaik A.R., 1997,
  %``The Identification of EGRET Sources with Flat-Spectrum Radio Sources,''
  ApJ, 481, 95
%arXiv:astro-ph/9612187.
  %%CITATION = ASTRO-PH 9612187;%%

\bibitem[\protect\citeauthoryear{von Montigny et al.}{1995}]{strong-emitters1}
von Montigny C. et al., 1995,
  %``High-energy gamma-ray emission from active galaxies: EGRET observations and
  %their implications,''
  ApJ, 440, 525
  %%CITATION = ASJOA,440,525;%%

\bibitem[\protect\citeauthoryear{Padovani \&
Giommi}{1995}]{PadovaniGiommi95}
Padovani P., Giommi P., 1995,
  %``The Connection between X-ray- and Radio-Selected BL Lacertae Objects,''
  ApJ, 444, 567
 % [arXiv:astro-ph/9412073];
  %%CITATION = ASTRO-PH 9412073;%%

\bibitem[\protect\citeauthoryear{Thompson, Bertsch
\& O'Neal}{2004}]{EGRET:highest}
  Thompson D.J., Bertsch D.L., O'Neal R.H., 2004,
  %``The highest-energy photons seen by the Energetic Gamma Ray Experiment
  %Telescope (EGRET) on the Compton Gamma Ray Observatory,''
  preprint (astro-ph/0412376); the catalog is available at {\tt
ftp//:} {\tt gamma.gsfc.nasa.gov/pub/VHE\_events/VHE\_events/} {\tt VHE\_events.tbl}
  %%CITATION = ASTRO-PH 0412376;%%

\bibitem[\protect\citeauthoryear{Thompson et al.}{1993}]{EGRET-resolution}
Thompson D.J. et al., 1993, ApJS, 86, 629

\bibitem[\protect\citeauthoryear{Tinyakov \& Tkachev}{2001}]{BL1}
 Tinyakov P.G., Tkachev I.I., 2001,
%``BL Lacertae are sources of the observed ultra-high energy cosmic rays,''
JETP Lett., 74, 445
%%CITATION = ASTRO-PH 0102476;%%

\bibitem[\protect\citeauthoryear{Tinyakov \& Tkachev}{2002}]{BL:GMF}
Tinyakov P.G., Tkachev I.I., 2002,
%``Tracing protons through the galactic magnetic field: A clue for
%charge
%composition of ultrahigh-energy cosmic rays,''
Astropart.Phys., 18, 165
%%CITATION = ASTRO-PH 0111305;%%

\bibitem[\protect\citeauthoryear{Tinyakov \&
Tkachev}{2004}]{cuts-and-penalties}
Tinyakov P.G., Tkachev I.I., 2004,
  %``Cuts and penalties: Comment on 'Clustering of ultrahigh energy cosmic rays
  %and their sources',''
  Phys.\ Rev.\ D, 69, 128301
  %%CITATION = PHRVA,D69,128301;%%
% arXiv:astro-ph/0301336.
  %%CITATION = ASTRO-PH 0301336;%%

\bibitem[\protect\citeauthoryear{V\'eron-Cetty \& V\'eron}{2003}]{Veron2003}
V\'eron-Cetty M.P., and V\'eron P., 2003,
    A\&A, 412, 399

\bibitem[\protect\citeauthoryear{V\'eron-Cetty \& V\'eron}{2000}]{veron}
V\'eron-Cetty M.P., and V\'eron P., 2000,
A\&AR, 10, 81

\end{thebibliography}
\end{document}